\documentclass{JINST}
\usepackage{cite}

\title{Photodegradation Mechanisms of Tetraphenyl Butadiene Coatings for Liquid Argon Detectors}

\author{B.J.P. Jones$^a$\thanks{Corresponding Author}, J.K. VanGemert$^b$, J.M. Conrad$^a$, A. Pla-Dalmau$^b$.\\
\llap{$^a$}Massachusetts Institute of Technology,\\
  77 Massachusetts Avenue, Cambridge, MA 02139, United States of America\\
\llap{$^b$}Fermi National Accelerator Laboratory,\\
  PO Box 500, Batavia, IL 60510, United States of America\\
  E-mail: \email{bjpjones@mit.edu}}

\abstract{We report on studies of degradation mechanisms of tetraphenyl butadiene
(TPB) coatings of the type used in neutrino and dark matter liquid
argon experiments. Using gas chromatography coupled to mass spectrometry
 we have detected the ultraviolet-blocking impurity benzophenone (BP).
We monitored the drop in performance and increase of benzophenone concentration in
TPB plates with exposure to ultraviolet (UV) light, and demonstrate the correlation between
these two variables. Based on the presence and initially exponential increase in the concentration of benzophenone observed,
we propose that TPB degradation is a free radical-mediated photooxidation reaction,
which is subsequently confirmed by displaying delayed degradation
using a free radical inhibitor. Finally we show that the performance
of wavelength-shifting coatings of the type envisioned for the
LBNE experiment can be improved by 10-20\%, with significantly delayed
UV degradation, by using a 20\% admixture of 4-tert-Butylcatechol.}


\begin{document}

\section{The Use of TPB In Neutrino and Dark Matter Experiments}

Liquid argon is the active medium in many existing and forthcoming
particle physics detectors. Examples include liquid argon
time projection chambers (LArTPCs) for neutrino studies, such as the
ICARUS~\cite{Menegolli:2012jq}, ArgoNeuT~\cite{Palamara:2011jy}, MicroBooNE~\cite{Jones:2011ci}, and LBNE~\cite{Akiri:2011dv} experiments,
and low-background scintillation detectors for dark matter searches
such as the DEAP~\cite{Boulay:2012hq} and MiniCLEAN~\cite{Rielage:2012zz} experiments.

Being inexpensive and inert, argon makes an ideal active medium for
a LArTPC, where a long free-electron lifetime must be achieved. Argon
is also an excellent scintillator, producing several tens of thousands
of photons per MeV deposited. Further, high-purity argon is transparent
to its own scintillation light. These characteristics make argon
one of the favorite active media for low-background scintillation detectors. 

Many LArTPCs also take advantage of argon scintillation light, which brings
several benefits above simply recording charge deposits. Light collection
systems which detect the flash of light accompanying a neutrino event
can be used to acquire timing information about the event with a much
higher precision than is possible with charge measurements alone. This
timing information is an important background rejection tool for experiments
deployed in pulsed neutrino beams, and is of particular importance
for surface-based detectors such as MicroBooNE and LBNE, which both
plan to implement extensive light collection systems. A light collection
system in a LArTPC can cover a large detector volume with a relatively
small number of channels, making the formation of trigger logic more
straightforward than it would be using TPC wires, which typically
number in the tens of thousands. There are also opportunities to use
scintillation light for event reconstruction, especially given its
widely reported nontrivial time structure, which is used to determine
the local ionization density of keV scattering events in dark matter
detectors ~\cite{Kryczynski:2012yi}.

\begin{figure}[h]
\begin{centering}
\includegraphics[height=4cm]{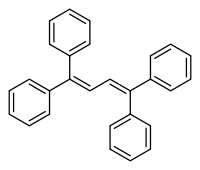}\qquad{}\includegraphics[height=4cm]{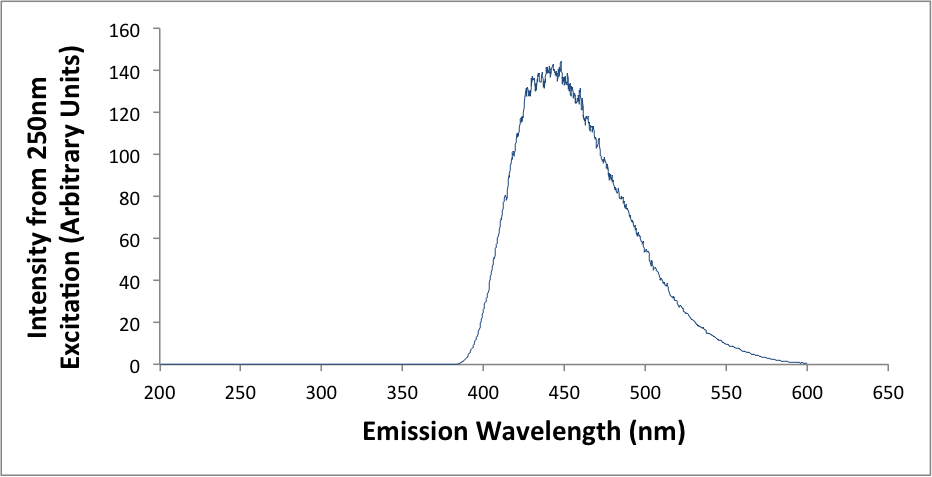}
\par\end{centering}

\caption{The chemical structure of TPB (left) and emission spectrum (right) \cite{Baptista:2012bf}
\label{fig:The-chemical-structure}}
\end{figure}

Scintillation light in liquid argon is produced at a wavelength of
128 nm. Argon is highly transparent at this wavelength, so the light
can be collected by detectors outside of the active region. However,
most optical detectors, including cryogenic photomultiplier tubes (PMTs),
are insensitive to 128 nm light. Tetraphenyl butadiene (TPB) is a fluorescent 
compound with the
chemical structure shown in Figure \ref{fig:The-chemical-structure}, which
has a high absorption coefficient at 128 nm and a visible
emission spectrum with a maximum around 450 nm.
Since this is a good match to the peak quantum efficiency of standard cryogenic
 PMTs, TPB is commonly used as a wavelength 
shifter in liquid argon scintillation detectors.
Thus, in order to make PMTs sensitive to argon scintillation light,
a coating of TPB is applied either to the PMT face, as in ICARUS, 
or to a separate plate which sits in front of
the PMT, which is the strategy favored by MicroBooNE, as well as MiniCLEAN.
A lightguide-based detector which transmits TPB-shifted light to
PMTs several meters from the interaction point is
also being developed, for the future LBNE detector \cite{Baptista:2012bf}.

There are several ways of applying TPB coatings, including evaporative
coating, dip coating and painting. Evaporative coatings of pure TPB
have the highest reported efficiency at 128 nm, with 1.3 visible photons
being produced for each one incident 128 nm photon~\cite{Gehman:2011xm}. These coatings
are fragile, however. In order to produce a more sturdy coating, TPB
is often mixed into a toluene / polystyrene solution and painted onto the surface. In this paper
we consider two types of films, which are similar to the coating planned
 for the MicroBooNE optical system and one coating being considered for the LBNE 
optical system, respectively. The MicroBooNE-type coating is
designed to give the highest light yield possible for a polystyrene-based 
coating, and uses a 1:1 TPB-to-polystyrene ratio, with 1 g of
each dissolved in 50 ml of toluene which is brush coated onto acrylic
plates. At this concentration, the TPB crystalizes out of solution
onto the plate surface as the toluene evaporates, leading to an opaque,
rough finish. This coating will be referred to as the ``rough''
coating.  The LBNE-type coating is designed to give a smooth optical surface 
for application on lightguides. For this coating, a 1:2 TPB-to-polystyrene 
ratio is used, with 0.5 g of TPB per 50 ml of toluene.
With this composition, the TPB stays in solution as the coating dries,
leading to a transparent polystyrene film with TPB suspended in the
polystyrene matrix. This coating will be referred to as the ``smooth''
coating. 

A third, polystyrene-free coating was prepared for GCMS studies, as polystyrene
solutions were found to be damaging to the chromatography column. 
This coating consists of 1g TPB per 50 ml toluene brush coated onto the plate surface. 
All coated plates used in this study were cut from acrylic sheets of thickness
1/8" and were of size 4 cm x 4 cm.  Half a milliliter of one of the aforementioned
solutions was dripped onto the plate surface and then spread and smoothed
with an acid brush. All plates were dried for at least 30 minutes in
a fume hood before use.

\section{Degradation Behavior of TPB \label{sec:Degradation-Behaviour-of}}

It has been previously reported that TPB coatings degrade when exposed
to UV light, both losing efficiency ~\cite{Chiu:2012ju} and turning yellow
in color ~\cite{Jerry:2010zj} as a function of exposure. This UV sensitivity is
an undesirable feature in light collection systems for large experiments,
since it leads to more difficult installation procedures and comes
with the risk of in situ degradation during the lifetime of the detector.
Gaining understanding of and seeking strategies to prevent such degradation
is therefore of significant interest in the field of argon scintillation
detection, and may be of value to a wider community of TPB users.

In this paper, we measure the wavelength-shifting 
performance of TPB coatings with a Hitachi F-4500 fluorescence
spectrophotometer, using an incident wavelength of 270 nm and measuring
the emitted intensity at the TPB emission peak. This strategy prevents
backgrounds from direct reflections or ambient light
from contributing significantly to the measured efficiency. Using a reference plate
kept in the dark we can correct for spectrophotometer drift, which leads
to changes of a few percent in the reported efficiency over the duration
of any study.

Two grades of TPB are commercially available: scintillation grade,
which has a purity specification of >99.9\%, and standard grade, which
has a purity specification of >99\%. It has been noted before that
the standard grade product is much more susceptible to discoloration,
turning bright yellow after a few hours of exposure to sunlight.
The scintillation grade product does not turn bright yellow, rather
an off-white color, and has a longer characteristic degradation time.

\begin{figure}[h]
\begin{centering}
\includegraphics[width=0.8\columnwidth]{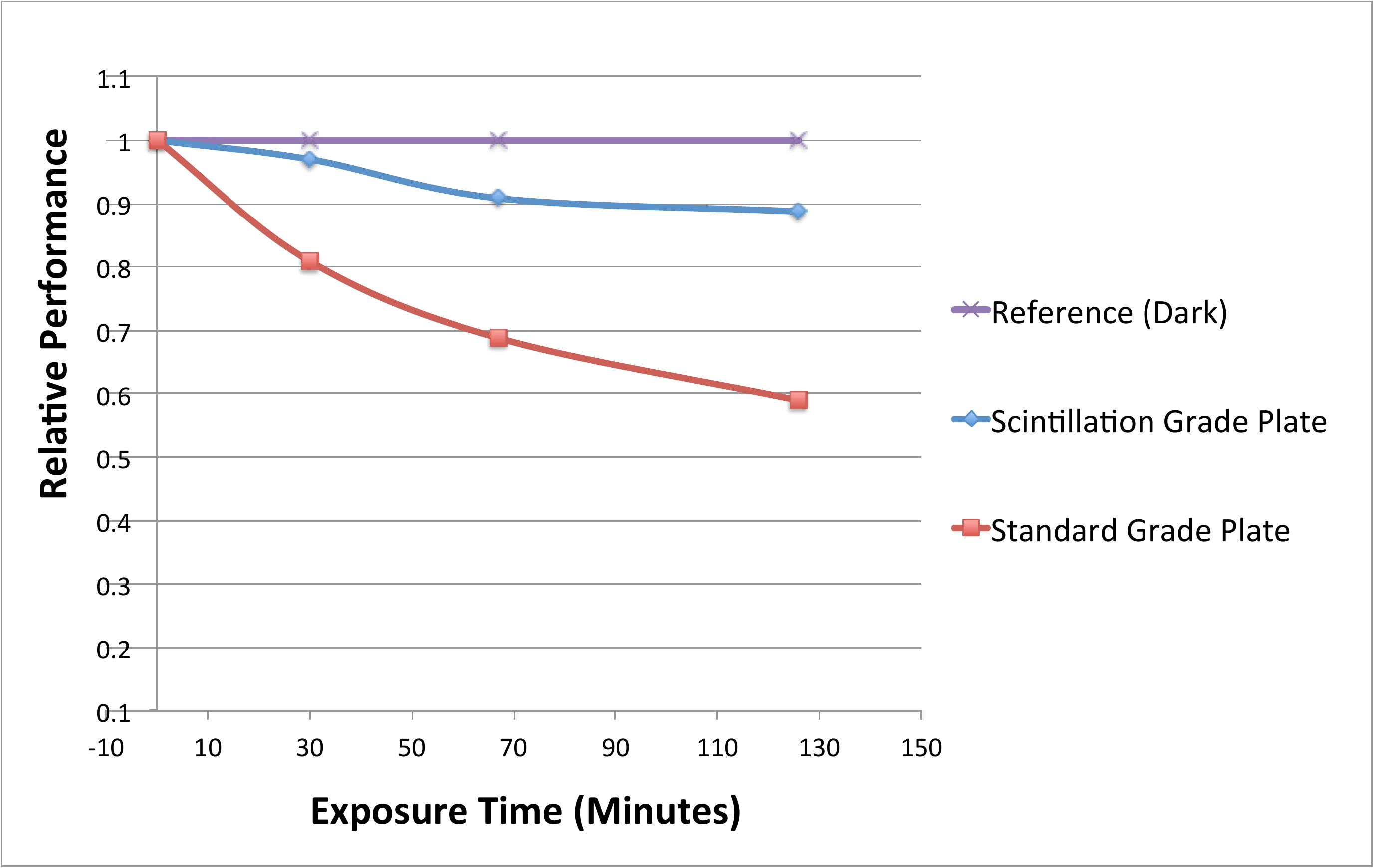}
\par\end{centering}

\caption{Degradation of standard and scintillation grade TPB plates with ``rough" coating, as a function of exposure to direct sunlight \label{fig:Degradation-of-a}}

\end{figure}

The relative efficiency of plates produced using the ``rough" coating
with standard and scintillation grade TPB as a function of exposure
to sunlight is shown in Figure \ref{fig:Degradation-of-a}. Both curves
are normalized to have their initial efficiency be equal to 1, and
each data point is corrected for spectrophotometer drift using the measured
efficiency of the reference plate. 

The intensity of the sunlight naturally varied over the course of
the study, so the scale on the horizontal axis should not be taken
to be linear. However by using sunlight we can ensure that both plates
received the same exposure, which has proved difficult to implement
with lamp=based studies, where slight changes to the geometry of the
setup or stability of the lamp are difficult to control. In this study, the standard
grade plate is seen to degrade significantly faster than the scintillation
grade plate. The observation that the standard grade sample both
loses performance and experiences discoloration at a higher rate than
the scintillation grade sample is suggestive that both of these features
are a consequence of initial impurities present in the sample.

\section{Identification of Impurities in TPB Powders With GCMS \label{sec:Identification-of-Impurities}}

\subsection{Use of GCMS to Identify and Quantify Impurities}

We used an Agilent 7820A Gas Chromatograph with a 5975 Mass Selective Detector (GCMS) to identify and quantify impurities in TPB powders and coatings.  This process involves two stages. First, different components of the sample are separated by their diffusion time in a gas chromatography (GC) column.  Then, each chromatographically separated component is fed into a mass spectrometer (MS).  The mass spectrum can be compared to a database of known spectra to elucidate a likely identity for the component.

After making preliminary mass-scan runs which identified impurities present in the sample, a custom GC method was created to accurately quantify their concentrations.  This method incorporated a temperature gradient of 120$^\circ$C to 250$^\circ$C, increasing at a rate of 5$^\circ$C per minute. This profile was chosen due to the different retention times exhibited by benzophenone and TPB.  Benzophenone would elute with the solvent at higher initial temperatures, whereas TPB would remain in the chromatography column at lower oven temperatures. The method was also set to maintain a constant flow rather than a constant pressure.  All injections were manual with a $ 5 \mu l$ syringe which was rinsed with solvent between each injection.

The MS was set to acquire data in selective ion mode (SIM) because of the sensitivity needed to detect the low benzophenone levels.  Each ion group was set to monitor the four most common ions for each compound.  Due the small size of the benzophenone peak compared to the TPB signal, in most cases a manual integration was required.

\subsection{Impurity GC Peaks in TPB Powders \label{sec:GCMSIdentify} }

\begin{figure}[h]
\begin{centering}
\includegraphics[width=0.9\columnwidth]{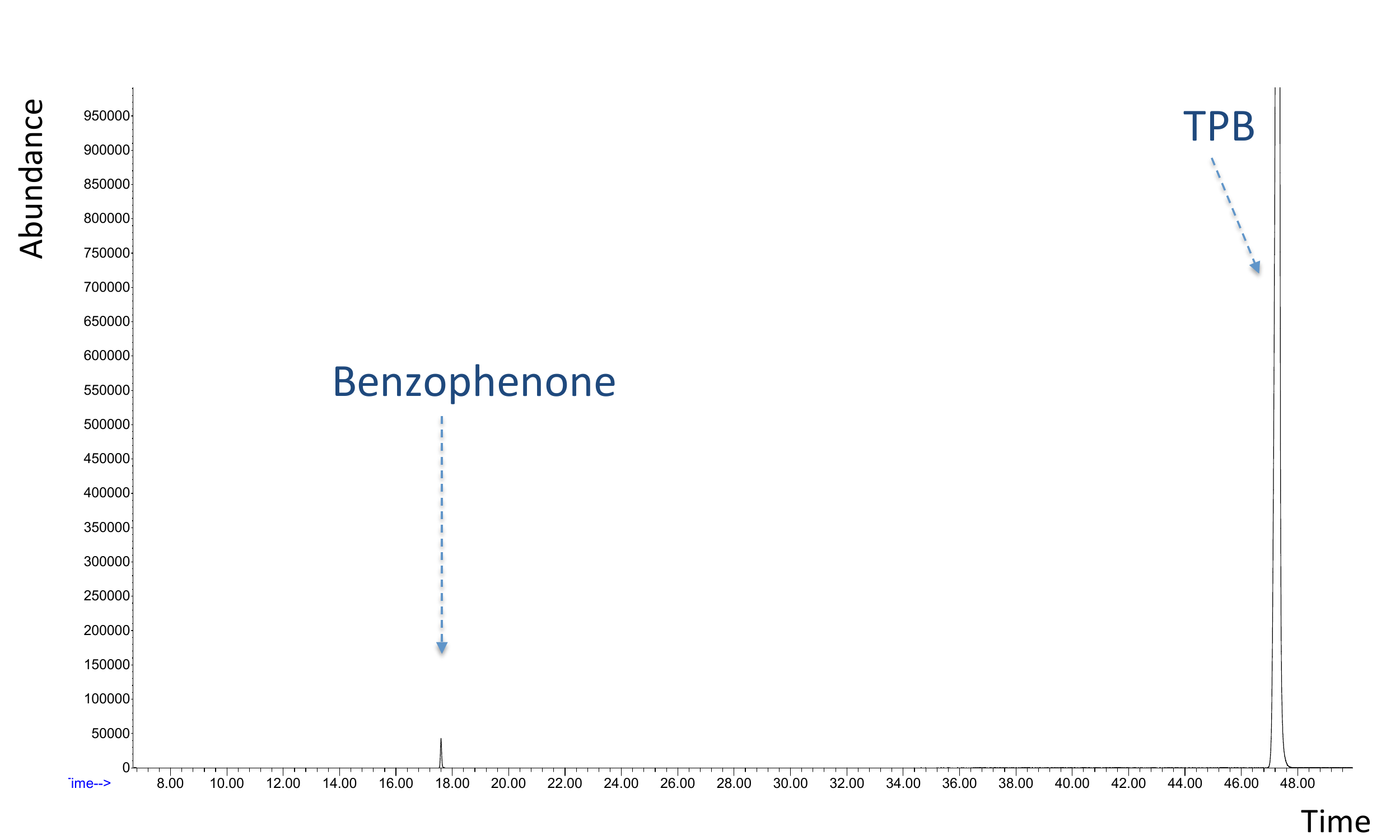}
\par\end{centering}

\caption{Chromatograph of unexposed standard grade TPB in toluene. \label{fig:Chromatograph-of-Standard}}

\end{figure}

An example chromatograph for standard 
grade TPB dissolved in toluene is shown in Figure
\ref{fig:Chromatograph-of-Standard}. The large peak at 47 minutes
is identified as TPB.  The peak at around
17.5 minutes is identified with a high confidence level as benzophenone,
which is a commercially available UV blocker and photoinitiator ~\cite{benzophenone}
and a likely photooxidation product of TPB.  The benzophenone concentration
was measured by comparing the BP to TPB peak areas in three powder
samples: unexposed standard grade, unexposed scintillation grade,
and standard grade TPB exposed to UV light to the point of yellowing. 
The benzophenone-to-TPB peak ratio from each are reported in Table \ref{tab:Benzophenone-to-TPB}

\begin{table}
\begin{centering}
\begin{tabular}{|c|c|}
\hline 
\bf{Sample} & \textbf{Benzophenone / TPB Peak Ratio}\tabularnewline
\hline 
\hline 
Scintillation Grade (Unexposed, White) & $4.13\times10^{-6}$\tabularnewline
\hline 
Standard Grade (Unexposed, White) & $8.43\times10^{-6}$\tabularnewline
\hline 
Standard Grade (Exposed, Yellow) & \textbf{$1.82\times10^{-5}$}\tabularnewline
\hline 
\end{tabular}
\par\end{centering}

\caption{Benzophenone-to-TPB peak area ratios in powder TPB samples. \label{tab:Benzophenone-to-TPB}}
\end{table}

When TPB is dissolved in toluene and applied as a plate coating, the
benzophenone concentration is seen to increase significantly, with the
 BP-to-TPB peak ratio increasing from order $10^{-4}\%$ in power samples, to 
around 0.01\% in samples taken from freshly coated plates.
This may be a result of the necessary light exposure involved in painting
and drying the plate, or of a reaction which occurs in solution. The
initial BP concentration varies between batches of plates, but sets
made from a common solution and subjected to a common storage procedure
appear to have consistent benzophenone concentrations to within 20\%.

\section{Monitoring Benzophenone Buildup in TPB Coatings}

The results of Section \ref{sec:Identification-of-Impurities} are
suggestive that the buildup of the benzophenone occurs when 
TPB is exposed to UV light. To test this hypothesis we prepared many identical
plates with scintillation grade TPB coatings. TPB was dissolved in 
toluene to make a solution of concentration 0.02 g/ml, 
and each plate received one coating made from
0.5 ml of this solution. As mentioned in Section 1, polystyrene was omitted
for this test because it was found that polystyrene solutions would not give consistent 
results and eventually lead to clogging of the GC column.

\begin{figure}[h]
\begin{centering}
\includegraphics[width=0.9\columnwidth]{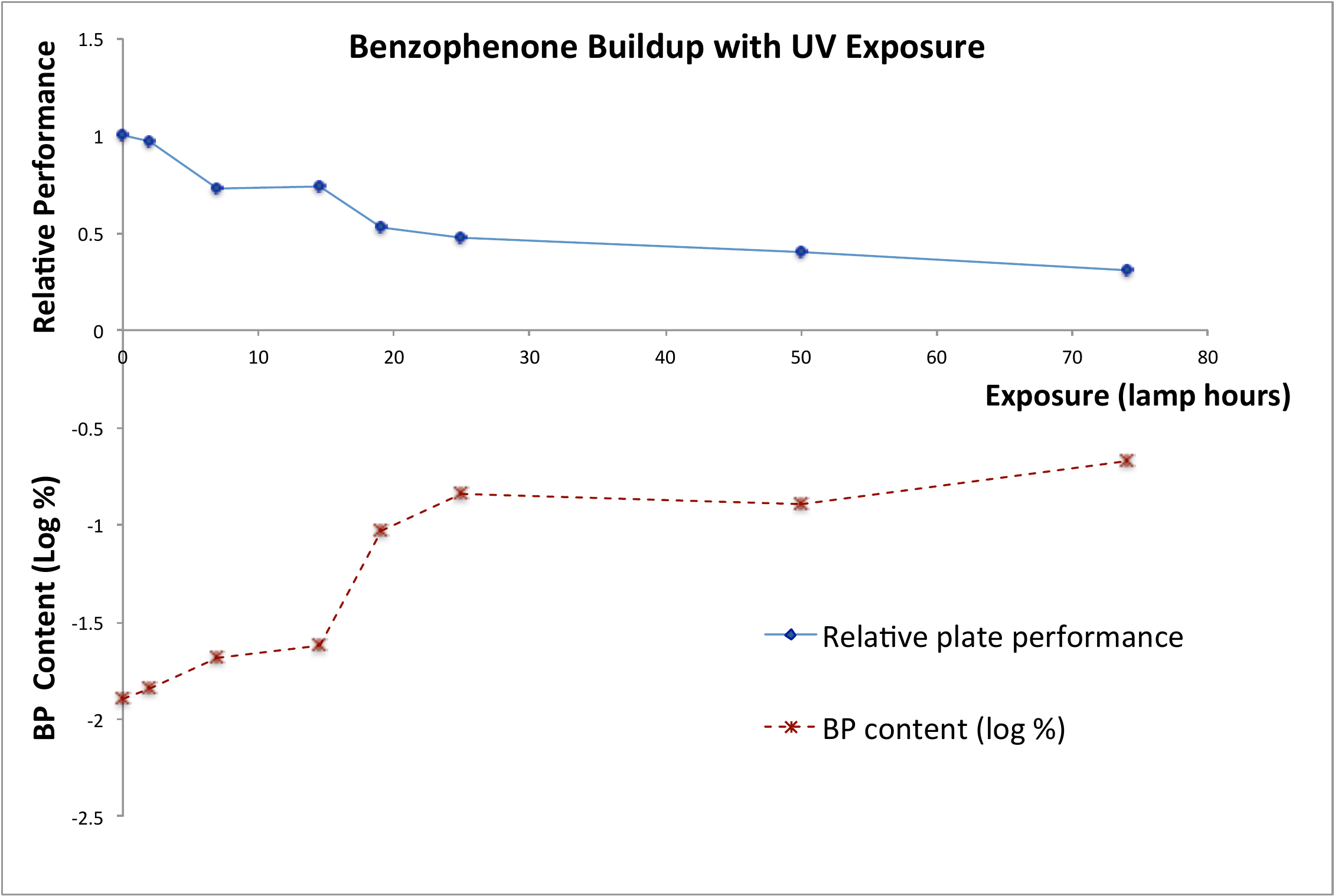}
\par\end{centering}

\caption{Degradation of plate performance and accumulation of benzophenone with exposure time \label{fig:Degradation-of-plate}}
\end{figure}

\begin{figure}[h]
\begin{centering}
\includegraphics[width=0.6\columnwidth]{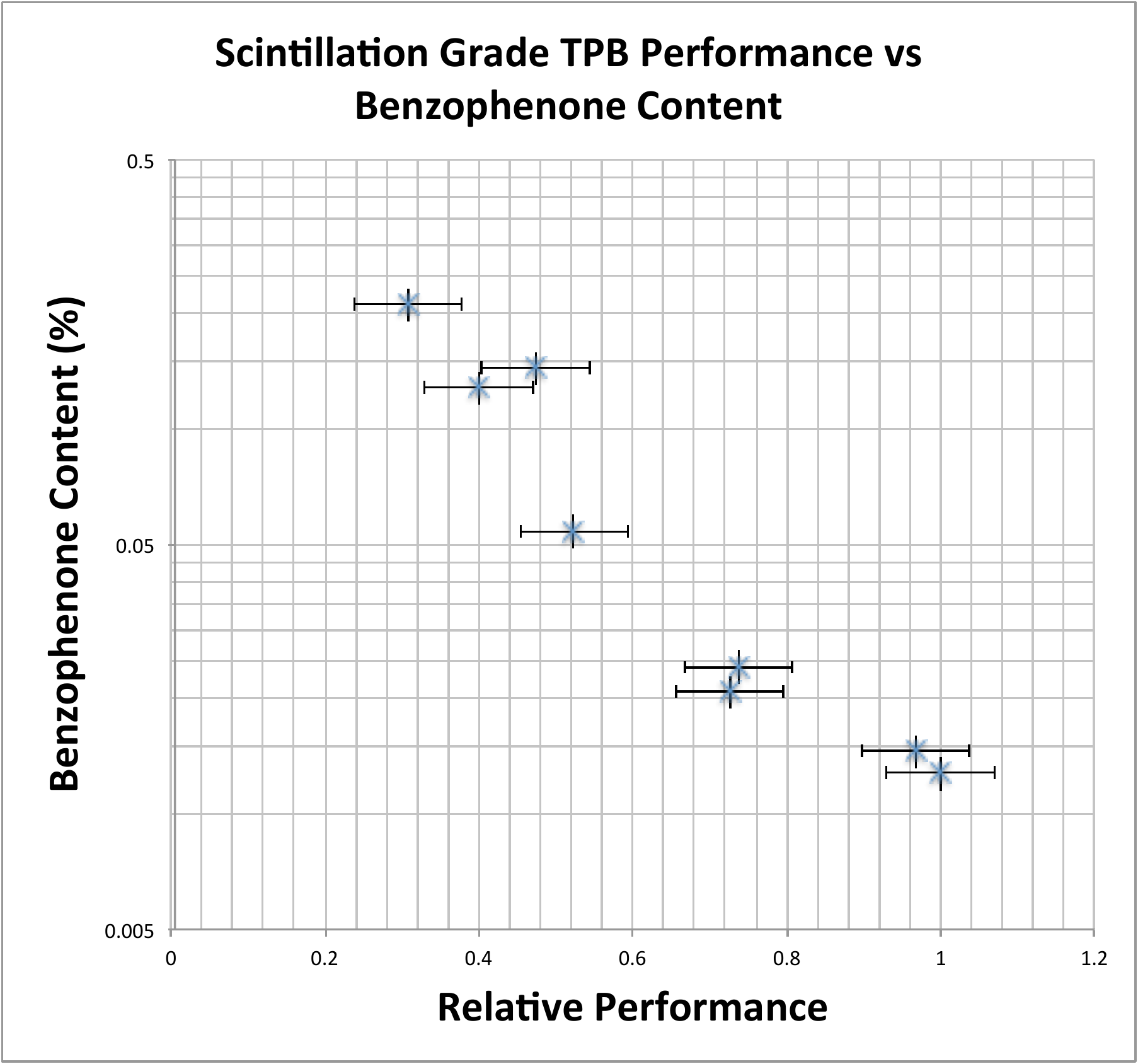}
\par\end{centering}

\caption{Performance vs benzophenone content (note log scale) \label{fig:Performance-vs-benzophenone}}
\end{figure}

Each plate was stored in the dark, wrapped in a mylar sheet, until
we were ready to begin exposing it to UV light. Immediately after removing 
each plate from the dark we measured the
initial wavelength shifting efficiency using the spectrophotometer procedure 
described in Section
\ref{sec:Degradation-Behaviour-of}. It was then exposed to a
365nm UV lamp for a number of hours,
and then the efficiency was measured again. The ratio of these two
measurements tells us the extent of the plate degradation. 
Next, the plate was placed into a
small bath of 20 ml toluene and sonicated
for five minutes in order to dissolve off the TPB
coating. Two $\mu l$ the solution was injected
into the spectrometer and the benzophenone / TPB peak ratio was recorded.
Using many plates we monitor the benzophenone content and plate performance
as a function of exposure time, which is shown in Figure \ref{fig:Degradation-of-plate}.
Since several plates were under the lamp simultaneously, the geometry
of the setup means that not all got exactly equal exposure. This leads
to some uncertainty in the total integrated UV exposure, which is evident in
the horizontal scatter observed in the data points of Figure \ref{fig:Degradation-of-plate}.
Plotting benzophenone content against plate performance loss bypasses
this issue and yields a very clear correlation, as is shown in Figure
\ref{fig:Performance-vs-benzophenone}. We assign a constant error
of 7\% in absolute performance to the spectrophotometer measurements, which
is determined by the spread of efficiencies measured for the unexposed
reference plate, and 10\% to the benzophenone / TPB peak ratio, which
is the approximate spread observed in repeated measurements of the
same solution.

\section{Benzophenone Spiking Studies, Active Coating Thickness and Discoloration}

We have observed the accumulation of sub-percent levels of benzophenone
in TPB coatings under exposure to UV light. Benzophenone is a commercially
available UV blocker and photoinitiator, so it is tempting to attribute the loss
of performance of TPB coatings to the partial absorption of incident light by 
the accumulated benzophenone, preventing it from being absorbed and re-emitted
by TPB.

\begin{figure}[h]
\begin{centering}
\includegraphics[width=0.9\columnwidth]{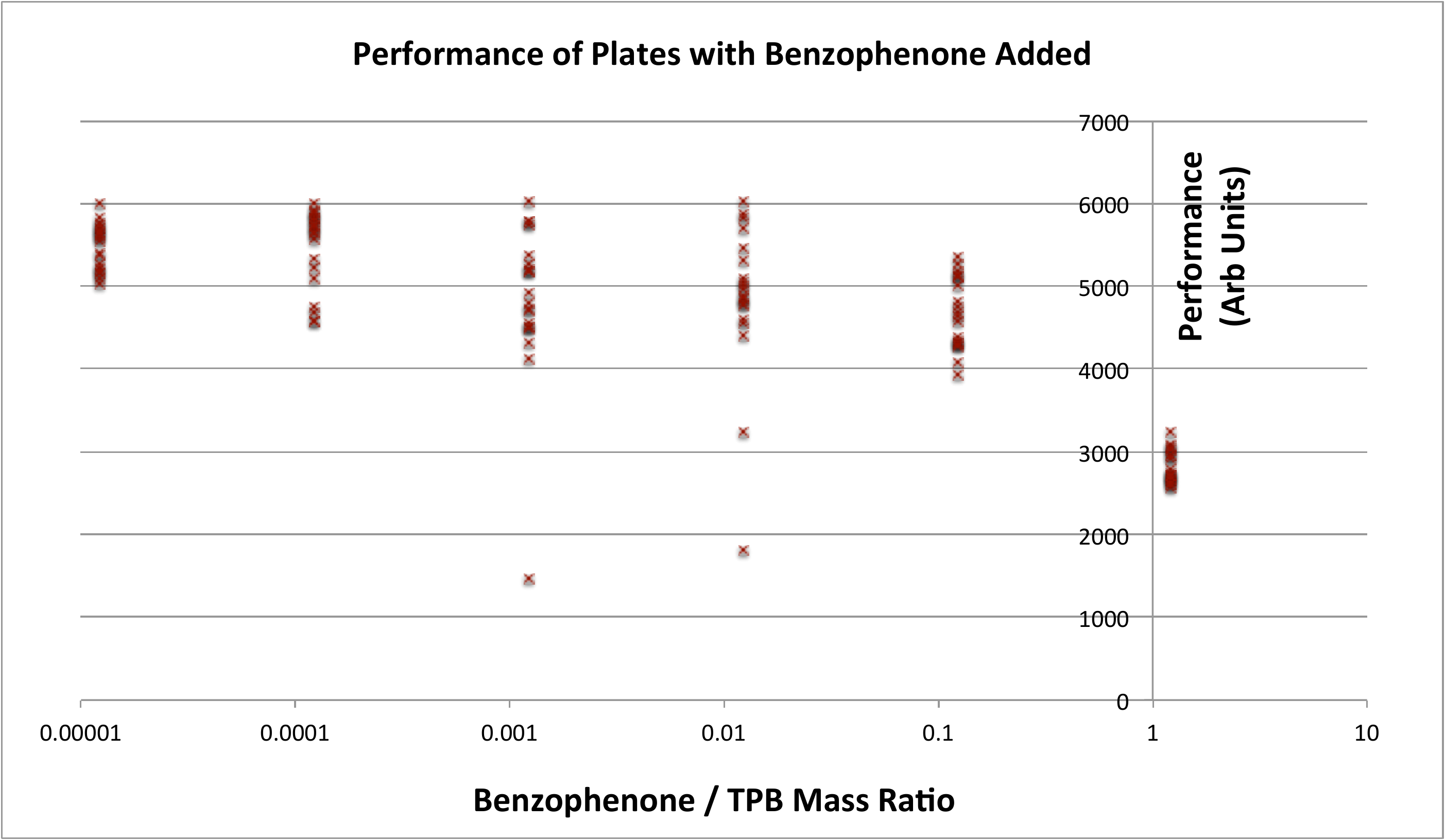}
\par\end{centering}

\caption{Performance of Benzophenone Spiked Plates \label{fig:Performance-of-Benzophenone}}
\end{figure}

In order to test the UV-blocking properties of benzophenone built up within 
the TPB layer, we prepared ``smooth" plates spiked with different benzophenone
concentrations and looked for a change in performance relative to
unspiked plates. In order to ensure that we were not misled by natural
plate-to-plate variations, we acquired high statistics by preparing
four plates at each concentration and measured the performance of
each plate at four different locations on its surface. The results
of this study are shown in Figure \ref{fig:Performance-of-Benzophenone}.

We note first that up to a benzophenone concentration of around 10\% by weight,
no significant loss of performance is observed, except for a few outlying
points which seem to indicate a tendency for a less uniform TPB distribution
within the coating when benzophenone is added. Only when there is twice 
as much BP as TPB do we see the levels of degradation observed in 
UV exposure studies. 

In TPB-coated plates, only the very top layer of
the coating under study is optically active, with all UV light being absorbed
in a thin surface layer.  In this case, the local benzophenone density at the plate
surface would be expected to be much higher than the global benzophenone 
concentration across the entire coating, since it has received a higher UV 
light exposure.  This produces an effective UV-blocking film of benzophenone which
prevents the underlying TPB from receiving the incident light.  

This type of skin effect is common in polymer photodegradation \cite{polymerbook}, and is to be expected in TPB, especially in light of recent studies which demonstrate that the efficiency of evaporative coatings is independent of their thickness \cite{thinfilmcoating}.

Our spiking results suggest
that the skin layer of degraded plates must have a composition of almost 100\% benzophenone,
and from this fact and the known density of polystyrene we can estimate the 
thickness of the optically active TPB layer to be of the order 100 nm, from a total
coating thickness of order 100 $\mu$m.

\begin{figure}[h]
\begin{centering} 
\includegraphics[height=4cm]{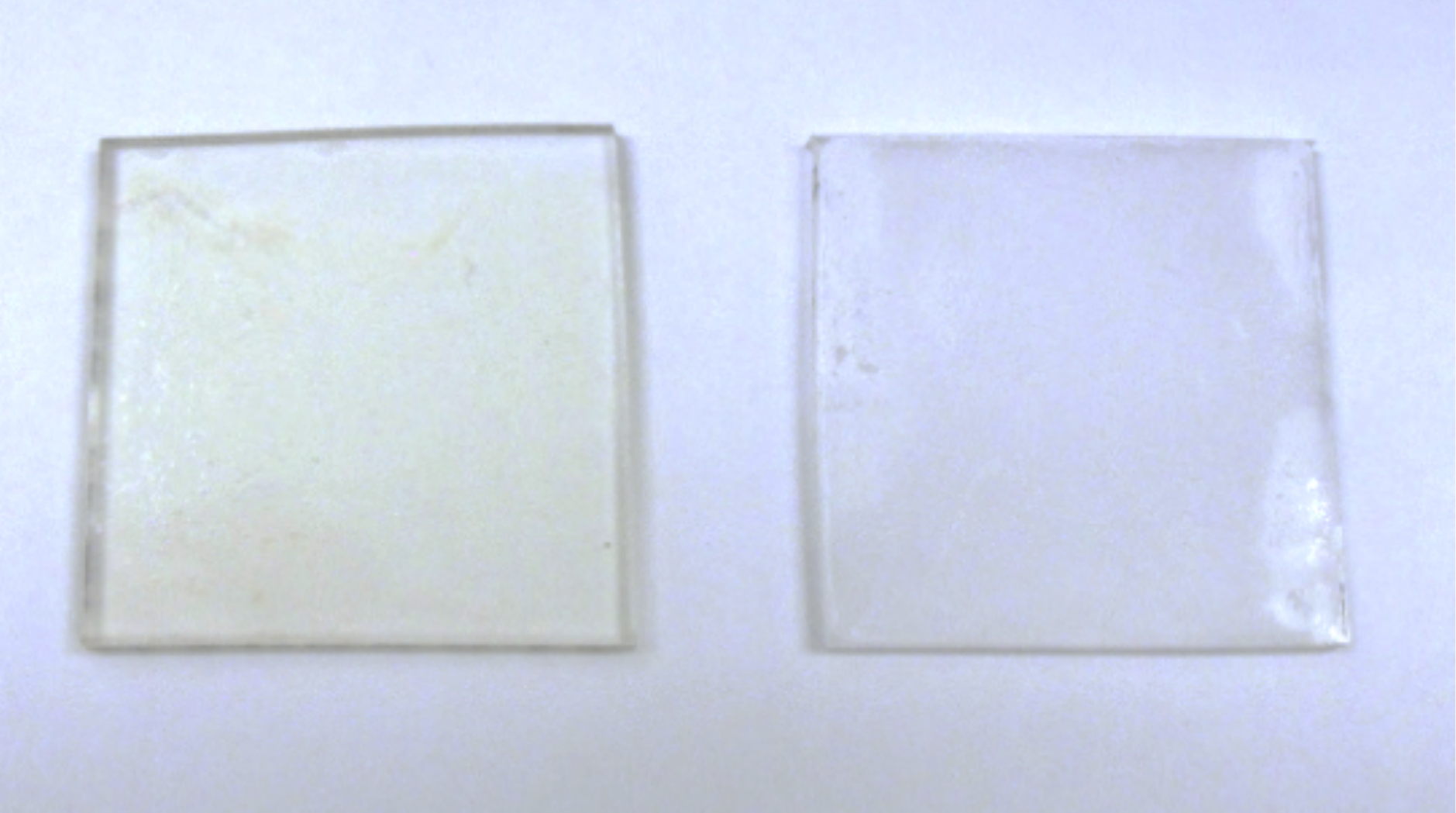}
\par\end{centering}

\caption{Similarly degraded plates with (left) and without (right) added benzophenone
\label{fig:Similarly-degraded-plates}}
\end{figure}

We also note an increased tendency for plates made with
added benzophenone to turn yellow in color under light exposure. A
photograph showing two similarly degraded plates, prepared with and
without added benzophenone, is shown in Figure \ref{fig:Similarly-degraded-plates}.  
We speculate that this effect may be due to the buildup of benzophenone derivatives, 
many of which are known to be both yellow in color and strongly UV-blocking \cite{MSDSSubBenzophenones}

\section{Chain Termination with Free Radical Inhibitors}
The amount of benzophenone accumulation displayed in Figure \ref{fig:Degradation-of-plate}
is initially approximately exponential as a function of light exposure.
This is suggestive of a free radical-mediated chain reaction, which is a
common mechanism of light-induced degradation, often encountered in polymer chemistry. \cite{PolymerChem}. Several
compounds exist which can be used to terminate such chain reactions
prematurely, and thus prevent the light-generated free radicals from
causing further degradation. To test whether this degradation is indeed
a free radical-mediated photooxidation, and to see if it is possible to
stabilize the coating, we investigated two such chain termination
compounds: 3,5-di-tert-butyl-4-hydroxytoluene (BHT) and 4-tert-Butylcatechol
(BC).

For both ``rough" and ``smooth" coatings, BHT was found to produce a less uniform
coating with regions of high and low TPB density. For this reason,
it was difficult to extract consistent results from coatings with
BHT admixtures. We are continuing to investigate methods of improving the
coating quality of BHT / TPB mixtures. 

BC admixtures did produce a homogeneous
coating. We even observed a slight tendency towards more TPB remaining
suspended in the polystyrene matrix rather than crystalizing on the
surface for the ``rough" plates. 

\begin{figure}[h]
\begin{centering}
\includegraphics[width=0.9\columnwidth]{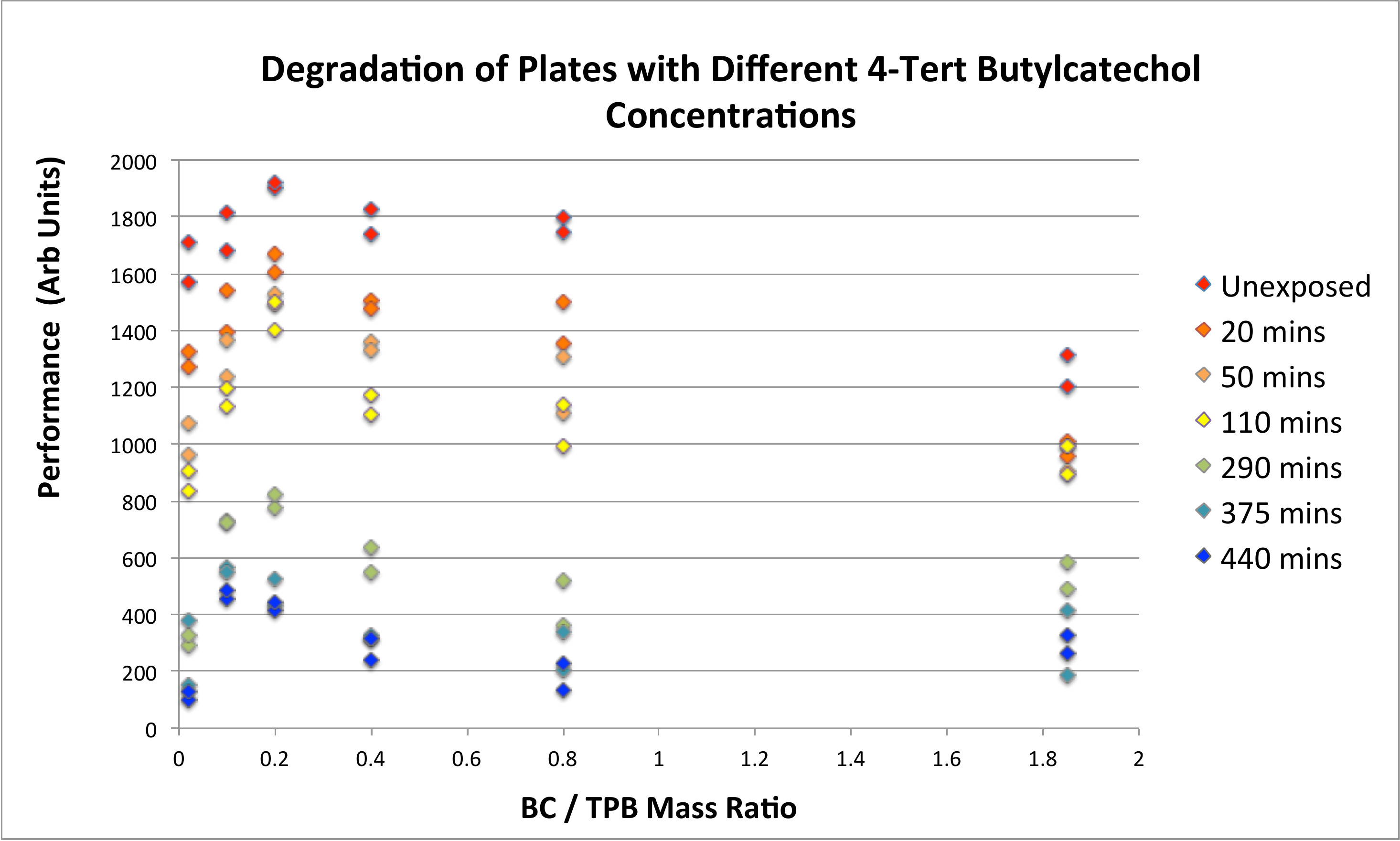}
\par\end{centering}

\caption{Degradation of plates with different BC / TPB mass ratios \label{fig:Degradation-of-plates}}

\end{figure}

We investigated plates with BC
/ TPB ratios between 0.1 ppm and 200\% by weight, and a stabilizing effect
was observed for mass ratios above 1\% BC / TPB. After establishing the relevant concentration
range, we prepared many plates with mass ratios between 1 and 200\% and monitored their degradation
over a few-hour period of simultaneous exposure to direct sunlight using the previously
discussed spectrophotometer procedure. Figure \ref{fig:Degradation-of-plates} shows the 
absolute efficiencies of plates prepared at various concentrations, as a function
of exposure time.  At each concentration point we used two identical, 
separately prepared plates, to illustrate the spread in coating performance. The plates used
were made with the ``smooth" coating. The results show that the addition
of around 20\% of BC appears not only to improve the initial performance
of the coating, but also to slow its degradation. This is a confirmation
that we are indeed seeing a free radical-mediated reaction. 

\begin{figure}[h]
\begin{centering}
\includegraphics[width=0.8\columnwidth]{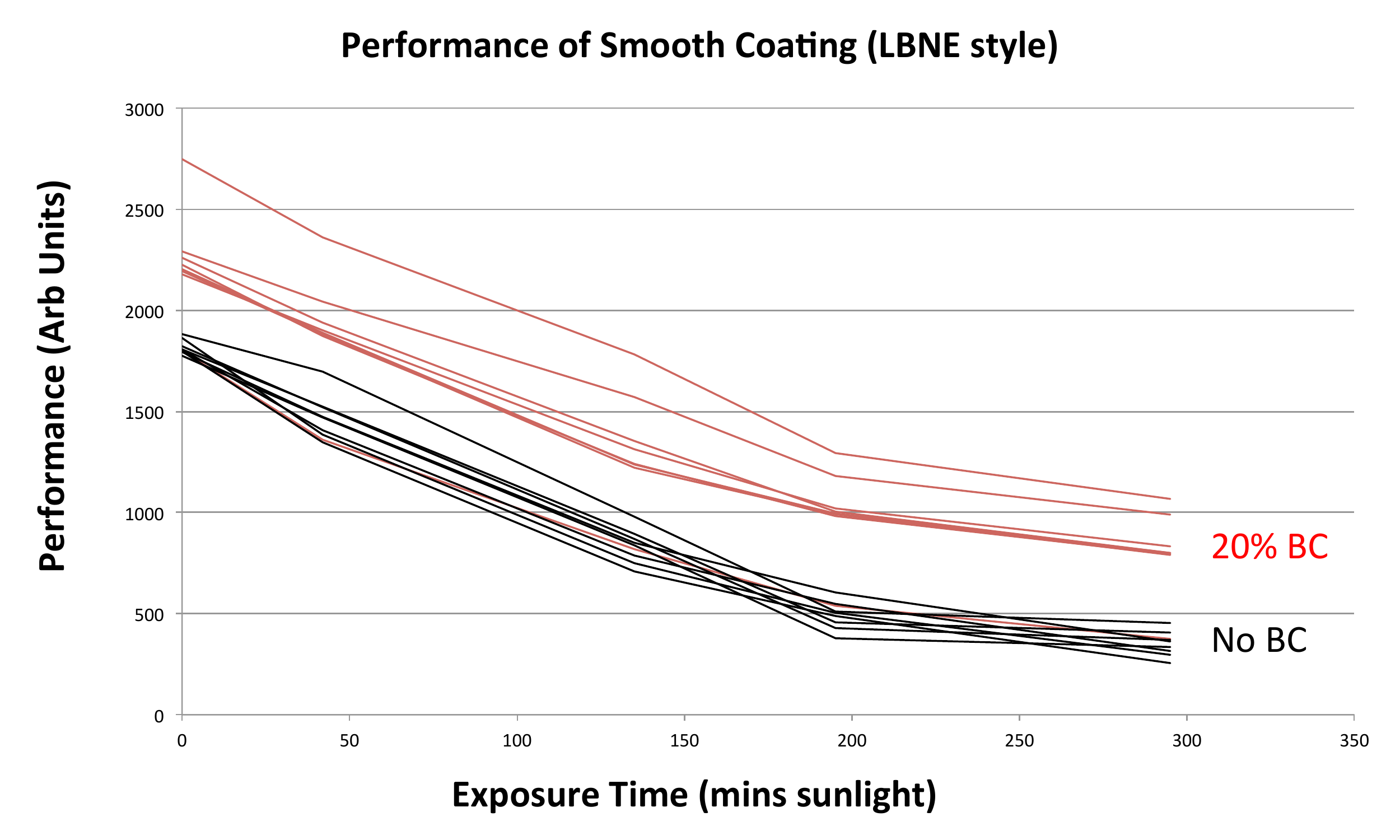}
\par\end{centering}

\medskip{}

\begin{centering}
\includegraphics[width=0.8\columnwidth]{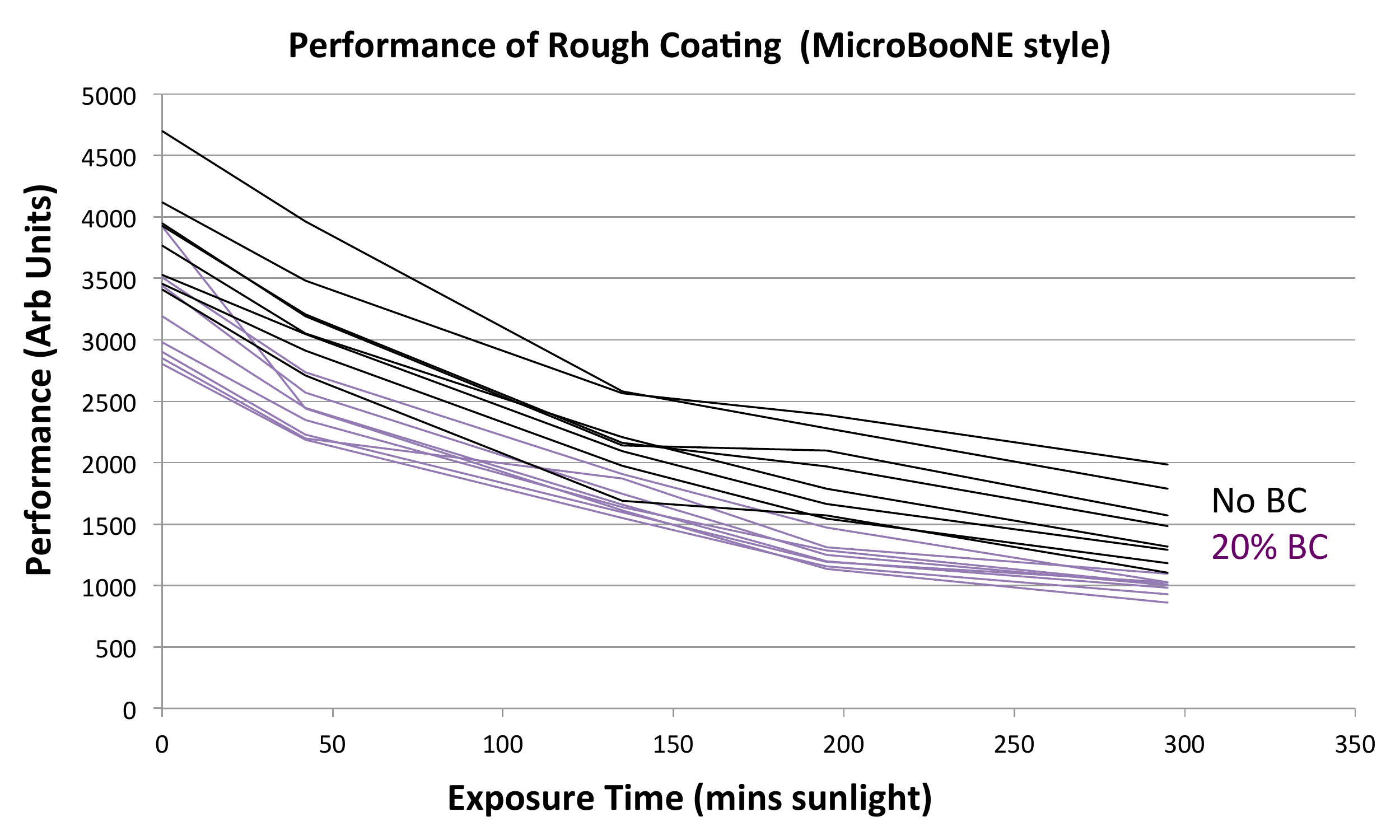}
\par\end{centering}

\caption{Study of 20\% vs 0\% BC plates, with both the "rough" and "smooth "style
coating \label{fig:Study-of-20=000025}}
\end{figure}

We tested that
this effect is repeatable by preparing many plates with 0\% and 20\%
BC, in both the ``rough" and ``smooth" styles, and monitoring their performance
as a function of sunlight exposure.
The data for this study are shown in Figure \ref{fig:Study-of-20=000025}.
We note that the ``smooth" plates reliably see a 10-20\%
improvement in initial light yield and an extended degradation time, such that
after 200 minutes of exposure, the stabilized coatings are \textasciitilde{}200\% as efficient
at the nonstabilized coatings.

The ``rough" plates are not improved.  This appears to be due
to the fact that less TPB crystalizes onto the surface when BC is
added, with more remaining suspended in the polystyrene layer. It may
be possible to turn this effect into an overall improvement for the
``rough" plates also, by increasing the quantity of TPB in the mixture to
restore the surface density to its value when no BC is added.
We plan to investigate this possibility in the near future.
Since the surface layer is responsible for most of the wavelength-shifting
capability of the ``rough" plates, and most of the BC remains in
the polystyrene matrix, we see little effect upon the coating lifetime
for this sample.

\section{Conclusions}

We have identified benzophenone as a compound produced in
TPB films which are exposed to ultraviolet light. Since benzophenone is a known ultraviolet
blocker and photoinitiator \cite{benzophenone}, it seems likely that its presence
is intimately related to the performance degradation observed in TPB
coatings.  Using GCMS and fluorescence measurements we have 
found a strong correlation between benzophenone concentration
and performance loss in TPB coatings with various UV exposures.

By preparing plates with BP / TPB mixtures we have shown that if benzophenone 
is the primary cause of performance loss, it must be accumulated within
a thin, optically active surface layer of the coating.  This type of skin 
effect is common in photochemistry \cite{polymerbook} and is
 supported by the results of other groups who have studied
TPB films of varying thicknesses \cite{thinfilmcoating}. 

The previously reported coating discoloration~\cite{Jerry:2010zj}  also appears to be related to the presence of
benzophenone, as plates spiked with additional benzophenone are observed
to turn significantly more yellow than TPB-only plates under exposure to similar light levels.

The initially exponential buildup of benzophenone under light exposure
is suggestive of a free radical-mediated TPB photooxidation reaction, which was tested by
exposing plates with an admixture of the free radical inhibitor 4-tert-Butylcatechol.
The delay in degradation provided by BC confirms that TPB degradation
is a free radical-mediated reaction. 

An admixture of 20\% BC by weight leads to an initial improvement of
the coating efficiency by 10-20\% and a delayed fluorescence degradation such
that after 200 minutes of sunlight exposure, the BC-protected coating
is more than 200\% as efficient as the unprotected coating. A program
to study other stabilizing compounds is underway. It is
possible that other free radical inhibitors may be found which can 
improve the lifetime of TPB coatings even more significantly.
\newpage
\acknowledgments

        We would like to thank Christie Chiu for her help in performing
preliminary studies related to this paper. The authors thank the National Science Foundation (NSF-PHY-084784) and Department Of Energy (DE-FG02-91ER40661).  This work was supported by the Fermi National Accelerator Laboratory, which is operated by the Fermi Research Alliance, LLC under Contract No. De-AC02-07CH11359 with the United States Department of Energy.

\newpage

\bibliography{TPBPaperTex}{}
\bibliographystyle{JHEP}

\end{document}